\documentclass[twocolumn,tightenlines,showpacs,aps]{revtex4}

\usepackage{graphicx}

\begin{document}

\title{Evidence of natural isotopic distribution from single-molecule SERS}

\author{P. G. Etchegoin, E. C. Le Ru, and M. Meyer}
\email{Pablo.Etchegoin@vuw.ac.nz; Eric.LeRu@vuw.ac.nz}

\affiliation{The MacDiarmid Institute for Advanced Materials and Nanotechnology,\\
School of Chemical and Physical Sciences, Victoria University of Wellington,\\
PO Box 600, Wellington, New Zealand.}

\date{\today}

\begin{abstract}
\noindent We report on the observation of the natural isotopic spread of carbon from single-molecule
Surface Enhanced Raman Spectroscopy (SM-SERS). 
By choosing a dye molecule with a very localized Raman active vibration in a cyano bond (C$\equiv$N triple bond), we observe (in a SERS colloidal liquid) a small fraction of SM-SERS events where the frequency of the cyano mode is softened and  in agreement with the effect of substituting $^{12}$C by the next most abundant $^{13}$C isotope. This example adds another demonstration of single molecule sensitivity in SERS through isotopic editing which is done, in this case, not by artificial isotopic editing but rather by nature itself. It also highlights SERS as a unique spectroscopic tool, capable of detecting an isotopic change in one atom of a single molecule. 
\end{abstract}

\maketitle

%\section*{Introduction}
As a spectroscopic technique, single-molecule Surface Enhanced Raman Scattering (SM-SERS) is now well established and accepted. A pr\'{e}cis of some of its most salient aspects has been recently provided in Ref.~\cite{PCCPfeature}. Arguably, one of the most refined and elegant demonstrations of Single Molecule Surface-Enhanced Raman Scattering (SM-SERS) is through the bi-analyte SERS technique \cite{2006LeRuJPCBBiASERS,2007GouletAnChem,2007MurakoshiJACS}, using isotopically edited molecules \cite{2007DieringerJACS,2008BlackiePCCP}. Isotopic editing is done purposely on specific moieties of standard SERS probes (like rhodamines \cite{2007DieringerJACS,2008BlackiePCCP}) to obtain molecules that have nominally identical chemical properties but distinguishable Raman features. 

Still, the typical atoms that constitute the structure of standard dyes used for SERS have their own natural isotopic spread. Accordingly, we can ask if the reverse logic applies: under SM-SERS conditions, is it possible to discern natural isotopologues? In this paper we show that --for specific cases-- the natural isotopic spread of carbon in organic dyes is indeed detectable.

%\section*{Background}

\begin{figure}[ht]
  \begin{center}
   \includegraphics[width=7cm]{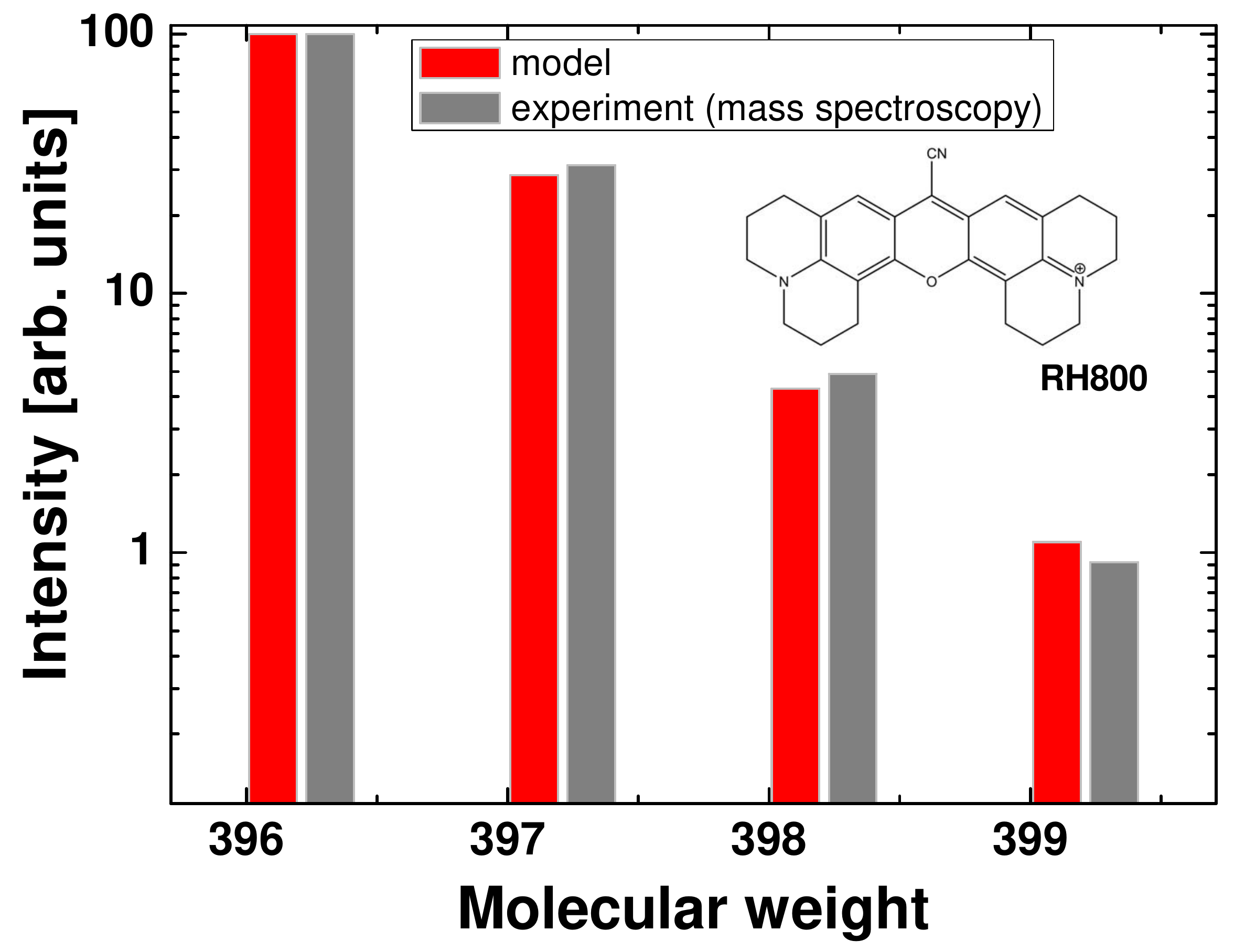}
   \caption{Experimental (gray) and calculated (red) isotopologues for rhodamine 800 (RH800, inset) produced by the natural isotopic spread of its constituent atoms. The experimental results have been obtained by high-resolution mass spectroscopy (MS). Approximately $\sim 20 \%$ of RH800 molecules have one more unit of mass because of the isotopic substitution of a $^{12}$C$\rightarrow$$^{13}$C in the structure (the rest corresponding to other isotopic substitutions). The number of molecules that have the $^{12}$C$\rightarrow$$^{13}$C substitution at the cyano bond (C$\equiv$N) is, however, much smaller ($\sim 1\%$) because it is only {\it one} site; i.e. it only depends on the natural isotopic ratio of carbon.}
\label{fig1}
\end{center}
\end{figure}

One of the obvious contributions to the isotopic distribution of organic molecules (of interest for SERS) is carbon. Despite a relatively small natural isotopic spread between $^{12}$C (98.9\%) and $^{13}$C (1.1\%) \cite{IUPAC}, carbon still plays a decisive role in the isotopic spread of organic molecules due to it preponderance with respect to other atomic species. Hydrogen is, of course, very abundant too in typical organic dyes, but this is compensated by its much smaller natural isotopic spread between hydrogen (99.985\%) and deuterium (0.015\%) \cite{IUPAC}. Let us study one particular example of a useful SERS probe: rhodamine 800 (RH800). The reason for choosing this particular molecule will become clearer later. The structure of RH800 is shown in the inset of Fig.~\ref{fig1}; it has the chemical formula: C$_{26}$H$_{26}$N$_3$O$^{+}$, and a molar mass of 396.21 g/mol when all the atoms are in their most abundant (lightest here) versions. Carbon and hydrogen are the most abundant atoms in the structure. Nitrogen and oxygen are not only a minority, but they also have smaller natural isotopic spreads with respect to carbon ($^{14}$N: 99.63\%; $^{15}$N: 0.37\%, and $^{16}$O: 99.75\%; $^{17}$O: 0.05\%; $^{18}$O: 0.20\%) \cite{IUPAC}. It is quite clear that by having the largest abundance and the widest (relative) natural isotopic spread (even when it is only $\sim 1\%$), it is enough for carbon to play the most predominant role.
In fact, the ``combinatorics'' of isotopic mass distributions are straightforward to calculate and can be compared to direct experimental measurements of the mass distribution of RH800 molecules, a common practice
in mass spectroscopy analysis. This is explicitly shown in Fig.~\ref{fig1}, obtained by standard high resolution mass spectroscopy on our sample. Note that mass spectrometry typically uses large (macroscopic) amounts of molecules to discern these differences.

These natural isotopic changes in mass result, in principle, in slight changes in vibrational frequencies. However, these cannot be resolved in most cases for a variety of reasons that include: $(i)$ the natural isotopic spread is very small, $(ii)$ natural isotopic substitutions occur at random places within the molecular structure, and not necessarily at places where they will considerably affect the eigenvectors and frequencies of Raman active vibrations, and $(iii)$ natural isotopic shifts are in many cases too small and hidden within the homogeneous linewidth of Raman peaks. In fact, a big fraction of the eigenvectors that produce Raman active modes in
medium-size or large molecules will be extended over many atoms \cite{book}, and the perturbation of a small mass change at a certain site will be relatively minor. From the standard theory of vibrations in molecules, the frequency of a given mode $\omega_{i}$ in a molecule is related to its {\it reduced mass} $\mu_i$ through the proportionality $\omega_{i} \propto (\mu_{i})^{-\frac{1}{2}}$ \cite{book}. The reduced mass can be viewed as a
eigenvector-weighted mass-average that takes into account the relative participation of different atoms in the collective motion described by a particular eigenvector. A change of mass $\Delta m_{j}$ on the $j$-th atom results in a relative change of frequency $\omega_i$ given by:
$\Delta\omega_{i}^{j} = -{\omega_i}/({2 \mu_i})\Delta\mu_i^j$ with  $\Delta\mu_i^j = ({\partial \mu_i}/{\partial m_j})\Delta m_j$.
%\begin{equation}
%\Delta\omega_{i}^{j} = -\frac{\omega_i}{2 \mu_i}~\left(\frac{\partial \mu_i}{\partial m_j}\right)~\Delta m_j.
%\label{eq1}
%\end{equation}
In general, this change will be small, and within the intrinsic linewidth (homogeneous broadening) of the peak.

\begin{figure}[ht]
  \begin{center}
   \includegraphics[width=7cm]{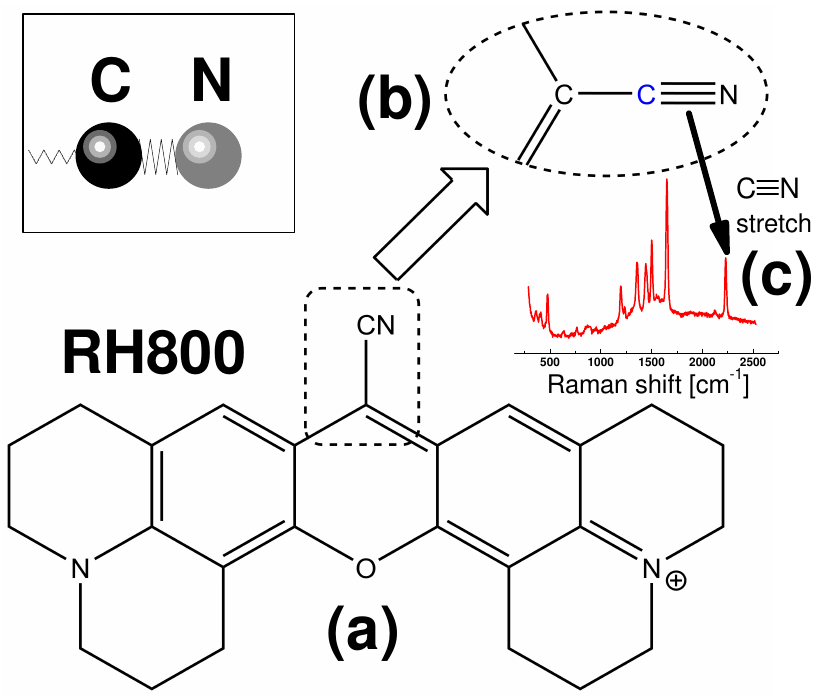}
   \caption{$(a)$ Molecular structure of RH800. In $(b)$ we show the immediate environment of the cyano bond (C$\equiv$N,) which forms a localized high frequency vibration resembling a ``dumbbell'' (inset). This vibration is Raman active, as shown in the spectrum in $(c)$, and appears at $\sim 2230\,{\rm cm}^{-1}$ in the so called ``Raman silent'' region (only triple bonds have Raman active vibrations in this region). Being a high frequency {\it localized} vibration, the cyano stretching mode is particularly susceptible to the isotopic substitution $^{12}$C$\rightarrow$$^{13}$C, resulting in a frequency softening of the mode by $\Delta\omega\sim 55\,{\rm cm}^{-1}$.}
\label{fig2}
\end{center}
\end{figure}

There are nevertheless exceptions, a family of which is associated with the existence of {\it localized vibrational modes} (for which $\partial \mu_i /\partial m_j$ is much larger).
RH800 has such a mode. The carbon-nitrogen triple bond (cyano group) in the structure of RH800, highlighted in Fig.~\ref{fig2}(a), is a strong bond that produces a relatively isolated and localized stretch vibration. The cyano group (highlighted in Fig.~\ref{fig2}(b)) produces a Raman active vibration at a frequency of $\omega\sim 2230\,{\rm cm}^{-1}$ for the most common isotopic combination of $^{12}$C$\equiv$$^{14}$N. This vibration is fairly localized to the stretch motion of carbon against nitrogen and occurs in the so-called ``Raman silent'' region; above the standard fingerprint region ($\sim 100-1700\,{\rm cm}^{-1}$) and below the hydrogen-stretching region ($\sim 2900-3100\,{\rm cm}^{-1}$) for typical organic molecules. The cyano stretch Raman-active mode is shown explicitly in the spectrum in Fig.~\ref{fig2}(c) for the specific case of RH800. Only triple bonds produce Raman active modes in this region. To a very good approximation, it is possible to think of the cyano bond as a ``dumbbell'' formed by the C and N atoms, coupled (through a ``softer'' spring) to the rest of the molecular structure. This is shown explicitly in the inset of Fig.~\ref{fig2}. By being a localized vibration with a high frequency, the C$\equiv$N is particularly susceptible to isotopic substitution and, in particular, to the most common $^{12}$C$\rightarrow$$^{13}$C replacement. The identification is helped by two facts: $(i)$ the shift also occurs in a region where there cannot be any overlap with other Raman active modes, thus facilitating its identification, and $(ii)$ the shift is large (because it is proportional to the frequency of the mode) and the peak is narrow enough to produce a new peak that cannot be confused as a small shift of the normal case (that could have other origins, like anharmonic/thermal effects). 

\squeezetable
\begin{table}[h]
%\begin{ruledtabular}
\begin{tabular}{|c|c|c|c|}
\hline
cyanobenzene & \scriptsize $\omega$ & \scriptsize $\Delta\omega$ & Abundance\\
C$\equiv$N stretch & [cm$^{-1}$] & [cm$^{-1}$] & \\
\hline
~~$^{12}$C$\equiv$$^{14}$N~~ &~~2332~~ &~~-~~ &~~98.53\,\%~~\\ \hline
~~$^{13}$C$\equiv$$^{14}$N~~ &~~2277~~ &~~55~~ &~~1.1\,\%~~\\ \hline
~~$^{12}$C$\equiv$$^{15}$N~~ &~~2303~~ &~~29~~ &~~0.37\,\%~~\\ \hline
~~$^{13}$C$\equiv$$^{15}$N~~ &~~2247~~ &~~85~~ &~~0.004\,\%~~\\ \hline
\end{tabular}
%\end{ruledtabular}
    \caption{\label{table} Carbon-nitrogen (triple bond) stretch for different isotopic combinations in cyanobenzene calculated by Density Functional Theory (DFT). The most likely case observed experimentally is the bond $^{12}$C$\equiv$$^{14}$N, followed by $\sim 1\%$ cases of $^{13}$C$\equiv$$^{14}$N which introduces a (calculated) frequency softening of $\Delta\omega\sim 55\,{\rm cm}^{-1}$. See the supplementary information for more details on the DFT calculations.}
\end{table}

Considering initially the most abundant case: $^{12}$C$\equiv$$^{14}$N, the reduced mass of the isolated ``dumbbell'' changes by  $\Delta \mu \sim 0.28$ for $^{12}$C$\rightarrow$$^{13}$C. . Hence, the predicted frequency change in the cyano frequency $\omega_{{\rm C}\equiv{\rm N}} \sim 2230\,{\rm cm}^{-1}$ is $\Delta \omega_{{\rm C}\equiv{\rm N}}\sim 50\,{\rm cm}^{-1}$. More accurate values that consider the link of the ``dumbbell'' to the rest of the structure can be obtained by direct calculations with Density Functional Theory (DFT). One advantage of the the cyano bond is that, being a localized vibration, it can be modeled with any molecule that resembles the local chemical environment of the bond. For example, we can study the cyano bond dynamics in RH800 with calculations on the much smaller molecule cyanobenzene (C$_{6}$H$_{5}$CN) with obvious computational advantages. Further details of the calculations are provided in the supplementary information. A summary of the results is provided in Table \ref{table}. A frequency softening of $\sim 55\,{\rm cm}^{-1}$ between the most abundant form of the bond ($^{12}$C$\equiv$$^{14}$N) and next most abundant one ($^{13}$C$\equiv$$^{14}$N) is predicted by DFT. Such a shift is much larger than the linewidth
of the corresponding Raman peak and therefore has to be easily observable in a small fraction of the SM-SERS spectra.

%\section*{Experiment \& Discussion}

\begin{figure}[ht]
  \begin{center}
   \includegraphics[width=7cm]{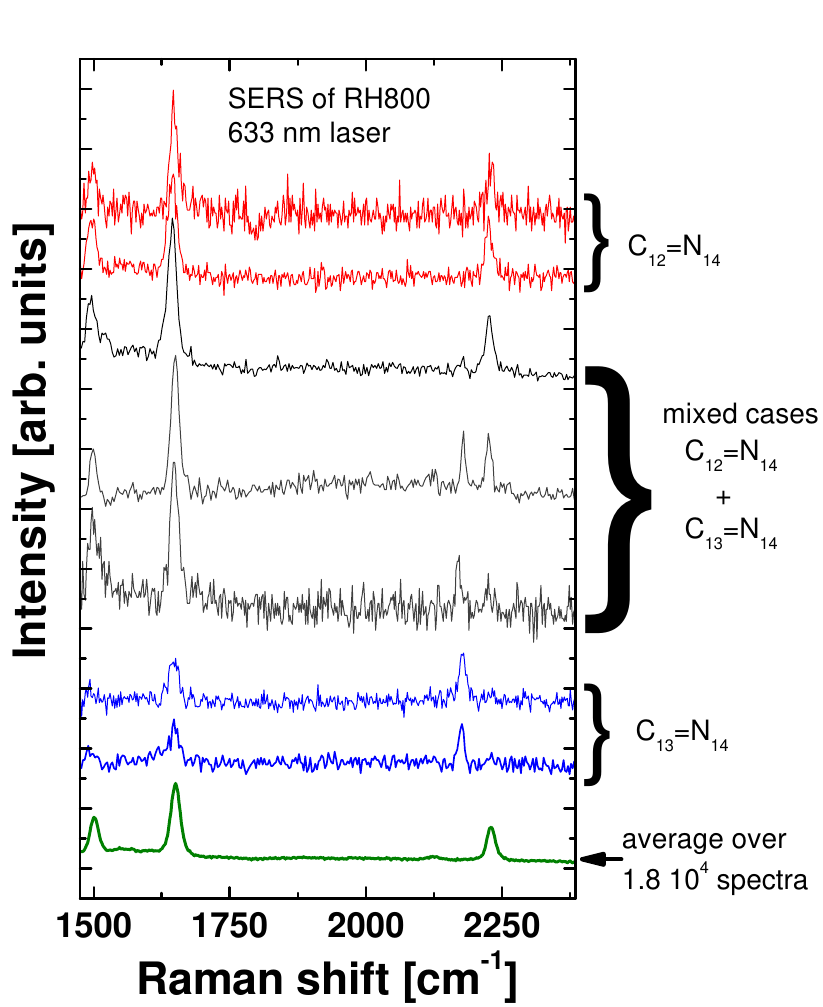}
   \caption{SERS spectra of RH800 showing from top to bottom: $(i)$ the normal spectroscopic feature of the Raman mode of the $^{12}$C$\equiv$$^{14}$N-cyano bond at $\sim 2230\,{\rm cm}^{-1}$ (top 2 spectra, red). $(ii)$ Mixed cases (next 3, black) where both the $^{12}$C$\equiv$$^{14}$N and $^{13}$C$\equiv$$^{14}$N-cyano bonds can be seen. These mixed spectra appear in the standard bi-analyte SERS method and we show three cases with larger, equivalent, or smaller relative intensities of one peak with respect to the other. $(iii)$ Single-molecule SERS events for the $^{13}$C$\equiv$$^{14}$N-cyano bond (next 2, blue). We also show the average spectrum (bottom, green) over the 18000 obtained spectra. Note that the number of cases where the $^{13}$C$\equiv$$^{14}$N-cyano bond signal is observed is very small, as evidenced by the negligible effect on the average spectrum. We show the spectra in a range that includes the $\sim 1650\,{\rm cm}^{-1}$ mode of RH800, which remains unchanged in all cases (as expected).}
\label{fig3}
\end{center}
\end{figure}

SERS spectra have been collected in immersion for RH800 in a silver (citrate-reduced) Lee \& Meisel colloid \cite{1982LeeJPC} at 10\,mM KCl. The experimental conditions were identical to those reported in our recent study of isotopologues of rhodamine in Ref. \cite{2008BlackiePCCP}. Spectra were obtained with a $\times 100$ objective indexed-matched to water and the 633\,nm line of a HeNe laser. Data were collected by a Jobin-Yvon LabRam spectrometer equipped with a notch filter and a nitrogen-cooled CCD. The dye concentration in the colloidal solution was 10\,nM and 18000 spectra were taken with 0.2\,sec integration time (with a 1\,sec dwell time in between spectra, to ensure statistical independence). We know from previous characterizations of our system for similar experiments with the bi-analyte method \cite{2008BlackiePCCP} that we are in the conditions where many of the spectra are single molecule in nature, with a fraction of them having contributions from more than one molecule.

Once the spectra are taken, we need to find cases where the isotopically modified cyano bond is present. This implies looking for a ``minority'' of cases that will not contribute much to the average signal. We have done this using Principal Component Analysis (PCA) (as explained in the supplementary information), but this is not necessary and any method should render equivalent results. The big advantage of PCA is that it reduces the search for the appropriate spectra to a matter of seconds; a task that becomes very tedious otherwise when 18000 spectra need to be analyzed.

Figure~\ref{fig3} shows representative spectra of the different cases of interest here. The vast majority of events show the cyano bond Raman peak at $\sim 2230\,{\rm cm}^{-1}$. A closer look, however, reveals situations with mixed signals at the corresponding frequencies expected for $^{12}$C$\equiv$$^{14}$N and $^{13}$C$\equiv$$^{14}$N. These are equivalent to the mixed signals that appear in the standard bi-analyte method. The experimental frequency softening is $\sim 56\,{\rm cm}^{-1}$, in excellent agreement with the DFT predictions (Table \ref{table}). We chose three spectra in Fig.~\ref{fig3} showing a larger, equivalent, or smaller intensity of the the $^{12}$C$\equiv$$^{14}$N-peak with respect to that for $^{13}$C$\equiv$$^{14}$N. Finally, we show cases where the signal comes {\it only} from the isotopically modified cyano bond ($^{13}$C$\equiv$$^{14}$N).
The average SERS spectrum is also shown for direct comparison.
It is interesting to note the small feature at $\sim 2120\,{\rm cm}^{-1}$ with
an integrated intensity $\approx 10$ times less than the cyano bond peak.
This is assigned to an overtone band and has no relation to the isotopic spread. It does however give an idea
of how small the $^{13}$C$\equiv$$^{14}$N Raman peak should be in the average signal: $\approx 10$ times smaller
than this $\sim 2120\,{\rm cm}^{-1}$ overtone band, i.e. clearly undetectable in the average spectrum under
the present conditions.
This is in stark contrast with the SM-SERS spectra, where the ($^{13}$C$\equiv$$^{14}$N)-cyano bond signal can completely dominate the spectra. The single-molecule nature of the signals here enable us to access information,
such as the isotope-induced shift, that are clearly washed out in the ensemble average.

% The situation here is, therefore, equivalent to a bi-analyte SERS experiment but with a very ``skewed'' distribution of molecules (98.9\% against 1.1\%); a situation that would normally be avoided in the standard version of the technique. The whole idea behind the bi-analyte technique is that it acts as a {\it contrast} method, in the statistical sense, and this is hindered for the most concentrated molecule if the concentration ratio is very skewed. But the uneven ratio here has been chosen by nature through the natural isotopic composition and, therefore, it provides us with a good way to identify single molecule cases of the least concentrated species. With this proviso in mind, all the standard conclusions for bi-analyte SERS experiments follow from here.
%

%\section*{Conclusion}

The results of this paper provide inderectly another demonstration of SM-SERS sensitivity to the ones already known \cite{PCCPfeature,2008PieczonkaCSR,Pettinger2008}. %It is paradoxical in a way that the field has gone through a decade of perfecting more elaborate proofs of single molecule sensitivity in SERS, including isotopic editing itself \cite{2007DieringerJACS,2008BlackiePCCP}, only to come back to an example that has been within reach all the time.
But with this aside, the main claim here is one step beyond another demonstration of SM-SERS sensitivity.
We have indeed shown a specific example of how single-molecule SERS can be used as tool for the observation of weak spectroscopic features that are otherwise washed out in ensemble averaged spectra.
 Other types of spectroscopy (like fluorescence) would be completely insensitive to a change of one unit in the isotopic mass of {\it one} atom, even if single molecule sensitivity is achieved. We believe the observation of naturally occurring single-molecule isotopic fluctuations in SERS is a pleasing demonstration of the consistency of SM-SERS phenomena in revealing subtle aspects of molecular spectroscopy that cannot be observed otherwise. The spectra in Fig. \ref{fig3} show, effectively, examples of single-molecule Raman spectra of molecules that differ from the others by {\it one unit} of atomic mass in {\it one} atom. Finally, it is also worth pointing out that the observation of natural isotopic spread with SM-SERS is not necessarily restricted to triple bonds. Options with chlorine-containing dyes could be particularly interesting to explore, for the natural isotopic spread of chlorine can produce a much more balanced list of naturally occurring isotopologues.

We are indebted to Sean Buchanan for help with the measurements. PGE and ECLR acknowledge partial support from the Royal Society of New Zealand (RSNZ) through a Marsden Grant.

\vspace{0.5cm}
Supplementary information available: details of DFT calculation  and PCA analysis of the SERS spectra.

\end{document}